# High performance new $MgB_2$ superconducting hollow wires

G.Giunchi[1], S.Ceresara[1], G.Ripamonti[1], A.DiZenobio[2], S.Rossi[2],
S.Chiarelli[3], M.Spadoni[3], R.Wesche[4], P.L.Bruzzone[4]

[1] EDISON SpA – Divisione Ricerca e Sviluppo, via U. Bassi 2  20159 Milano
[2] EUROPA METALLI SpA, via della Repubblica 857   55052 Fornaci di Barga(Lu)
[3] ENEA – Centro Ricerche Frascati – via E. Fermi 45   00040 Frascati (RM)
[4] EPFL-CRPP – Fusion Technology, CH-5232 Villigen –PSI

**Abstract**

$MgB_2$ hollow wires have been produced with a new technique which uses a conventional wire manufacturing process but is applied to composite billets containing the elemental B and Mg precursors in an appropriate shape. The technique has been applied to the manufacture of both monofilamentary and multifilamentary wires of several tens meters length. The superconducting transport properties of the $MgB_2$ hollow wires have been measured in magnetic field and in the temperature range from 4.2 to 30 K. Promising results are obtained, which indicate the possibility of application of these wires as superconductors in the temperature range of 15÷30 K and at medium-high values of magnetic field.

**Corresponding Author :**

Giovanni Giunchi
Via U. Bassi 2 – 20159 Milano (Italy)
Tel. +39 (02) 62223194
Fax  +39 (02) 62223074
e-mail : Giunchig@edison.it

## 1. Introduction

Short time after the discovery of the superconductivity in the $MgB_2$ at temperatures up to 39 K [1] several properties of this material favourable for applications were observed. Among the others: the persistence of the superconducting currents even in cold-worked but not sintered powders[2] and the almost complete magnetic field exclusion from the interior of highly dense polycrystalline bulk materials observed by magneto optical imaging [3]. Both characteristics indicated that the grain boundaries of the $MgB_2$ do not act as barriers for the superconducting currents as found in the high-Tc superconducting cuprates. This property opened the way to an easy manufacturing of the $MgB_2$ based superconductors.

From the beginning of the $MgB_2$ rush a large effort is dedicated in many labs to the manufacturing of wires or tapes from this material. Most of the attempts are based on the Powder In Tube (PIT) technology in its two major variants : a) the processing of $MgB_2$ powders ( ex situ) ; b) the processing of a mixture of Boron and Magnesium elemental powders (in situ)[4-8]. For this technology the best results are obtained when high stresses to the tape are applied, appropriate thermal treatments and fine powders are used. As a general rule it seems that the ex situ method provides better transport characteristics in high magnetic fields , but lower current density in low or self field. This behaviour can be explained in term of a more regular crystal growth of the in situ process , combined with a coarser grain morphology and a smaller amount of impurities. On the contrary the ex situ material, with reduced grain size and a high impurity level, presents better pinning features.

A new wire manufacturing technology, which combines features of the PIT process with the use of a reactive liquid infiltration, is here presented. According to this technology, hollow rounded wires are obtainable, with substantial better superconducting characteristics than the actual in situ wires and comparable with the best ex situ products.

The technology benefits of two peculiar characteristics: a) the cold-working of the Magnesium core rods, allows to obtain surprisingly large area reductions when the rods are dressed by fine grained powders, even without the application of intermediate annealing steps, b) after the reaction of the Mg with the B the last element remains fixed in the position where it was confined by the mechanical processing, giving rise to a corona shape morphology.

We present for the first time superconducting characteristics of these new wires as a function of temperature and applied magnetic field. The achieved critical current densities are high enough to be of interest for applications at moderately high magnetic fields up to 2-4 Tesla and at intermediate temperatures accessible by the use of cryocoolers.

## 2. Experimental details

The new wire manufacturing technique is based on our previous experience in preparing bulk $MgB_2$ bulk material by reactive liquid infiltration [9] .

The process includes the cold-working (by rod rolling, swaging and drawing) of a composite billet, that in the present case is made by soft steel internally lined with a niobium tube, filled with a coaxial internal cylindrical Magnesium rod and with fine grained Boron powders (Amorphous Boron, 95% pure, Starck AG), moderately compressed in the space between the Nb sheet and the Mg rod.

After this processing, the resulting wire reaches a cross section diameter of the order of the mm. After clamping its terminals with appropriate sealing tools the wire was annealed at temperatures between 750÷950 °C for 1÷3 hours leading to the formation of $MgB_2$.

Due to the reaction almost all the space previously occupied by the Magnesium rod is now void and the resulting $MgB_2$ is distributed, as a corona, on the inner wall of the wire. As for the bulk material obtained by the same infiltration technique, the resulting $MgB_2$ is very compact, dense and finely grained.

In the following we describe two typical preparations of hollow wires : a) a monofilamentary wire and b) a 7-filaments wire , both having a low carbon steel external case and the superconducting part laying inside a Nb sheet.

*2.1 Monofilamentary wire preparation* : the starting billet, 20 mm in diameter, contains the Mg rod and the B powders in the weight ratio of 1.55. At first it is rod rolled and swaged to a diameter of 2.3 mm, then it is drawn to a diameter of 1.5 mm and put in another steel case of 4 mm outer diameter and swaged and drawn up to a final outer diameter of 2.2 mm.

A short sample ( about 10 cm in length) taken from the obtained wire, has been sealed at both ends and annealed at 900°C for 3 hours.

Figure 1 shows an optical micrograph of a polished cross section of the superconducting monofilamentary wire.

*2.2 7-Filaments wire preparation* : the starting billet , 12 mm in diameter, contains the Mg rod and the B powders in the same ratio as in 2.1 and is swaged to a diameter of 3.66 mm . At this point the external steel was removed by an acid solution and the resulting rod, with its Nb sheath, was cut into 7 pieces and hexagonally assembled into another 12 mm external diameter steel tube. This new composite was further swaged to a final diameter of 2.5 mm. A short wire sample, 11 cm long, has been sealed at the extremities and annealed at 850°C for 30 minutes. Figure 2 shows the morphology of the cross section of the wires , before and after the annealing .

**3. Wires morphology**

From the cross section of the wires the area of the superconductor used for the computation of the critical current densities has been estimated. Other area values including the voids and the Nb sheet, together with their R ratio with the total area of the wires, can be of interest to estimate the area which is inherent to the present hollow wires morphology. We recall that the choice of the area of the external metallic case is somewhat arbitrary, provided that the stabilisation issues are considered and the mechanical strength to sustain the reaction pressure is guaranteed . The estimated area values of both types of hollow wires are reported in Table 1. The SEM analysis of the microstructure of the superconducting wires shows some typical features of these samples.

Figure 3 shows the interfacial region between the Nb sheet and the under laying $MgB_2$ material indicating a strong interaction of Nb and B probably resulting in the formation of Nb boride. In the monofilamentary wire the strong interaction is due to the high reaction temperature and to the long reaction time. The $MgB_2$ morphology , illustrated in the SEM analysis picture of Figure 4, shows a void free very dense grain packing, with a grain mean size of several micrometers.

The variable darkness of the grains reflects their different content of magnesium, which in our preparation is in excess with respect to the stoichiometric value: the darker grains being less magnesium contaminated in accord with the electron backscattering analysis that associate lighter colour for heavier elements. As reported for the bulk material prepared according to the same infiltration technology, the excess magnesium does not prevent the flowing of the superconductor currents throughout the grains[10]. The mean size of the pure grains of the $MgB_2$ part in the hollow wires is substantially smaller than that observed in our bulk samples in which the $MgB_2$ pure grains can extend for several tens of micrometers.

This feature is related to the granularity of the starting boron powders, that for the bulk sample is based on larger grain mean dimensions.

Looking for the grain size distribution of the 7-Filament wire, shown in Figure 5, the most evident result is a close similarity, in terms of grain size and morphology, to the previous wire, in spite of the very different reaction conditions.

On the contrary the 7-Filaments wire does not show an appreciable interaction between the $MgB_2$ and the Nb sheet, as evidenced in Figure 6.

## 4. Superconducting properties

The superconducting behaviour of the $MgB_2$ polycrystalline material obtained by the presented liquid infiltration technique has been described in detail for the bulk material[11-15]. Here we consider the properties deduced from the transport measurements of the above two hollow wires, in magnetic fields up to 10T and in the temperature range of 4.2 – 30 K. A temperature variable cryostat was used for the measurements at temperatures above 4.2 K.

### *4. 1 Critical current density*
The critical current densities were measured with the standard 4 point technique, defining the critical current by an 1 microV/cm electric field criterion. The Jc has been calculated using the superconductor area listed in Table 1.

The plot of the resulting Jc(B,T) values for the monofilamentary wire is shown in Figure 7.

Generally, the n values ($V \propto I^n$) describing the V(I) characteristics decrease with both increasing magnetic field and temperature. In our case the typical n values for fields lower than 1 T are higher than n=50. The plot of the resulting Jc (B,T) values for the 7-Filaments wire is reported in Figure 8

The n values measured in this wire range from 19 to 4 and depend inversely from the increase of the magnetic field and the temperature. In the presently considered interval of B we obtain n values that coincide with the values reported by Beneduce et al. [16] only at 10 T, at lower fields we obtain lower values, for example at 4.2K and 4T we have n=16 as compared to n=60 resulting for the sheathed tape of the ex situ type reported in [16].

Comparing the current densities of our two wires we found practically a similar behaviour in all the range of the examined fields and temperatures: in spite of differences in the morphology of the wires and also in the annealing conditions during reaction.

Comparing with other wire or tape manufacturing techniques [17,18] our record results of Jc at 4.2K of about 100 $A/mm^2$ at 6T and of about 4000 $A/mm^2$ at 2 T we are on the line of the best results published for the ex situ technique, but we are substantially better with respect to the in situ technique, to which our process can be assimilated.

*4.2 Irreversibility field*

To extract information about the temperature dependence of the irreversibility field, B*(T), which describes the limits of magnetic field/temperature zones practically accessible to applications of the superconductor in question, we have used the Kramer model scaling law of the Ic(B,T):

$$I_c(B,T) = \frac{C(T)}{B}\left(\frac{B}{B^*(T)}\right)^p \left(1 - \frac{B}{B^*(T)}\right)^q$$

where

$$B^*(T) = B^*(0)\left(1 - \left(\frac{T}{T_c}\right)^a\right)$$

and

$$C(T) = C_0\left(1 - \left(\frac{T}{T_c}\right)^b\right)^g$$

The scaling parameters are listed in Table 2.
The selection of p and q parameters considerably different from the conventional values of p=0.5 and q=2, is necessary to obtain an almost linear behavior in the Kramer's plot over the entire B field interval.
The $\alpha$ parameter, according to various flux lattice dynamic theories, assumes the value 2 for the lattice melting model[19] or the values 1.33 or 1.5 for a thermal activation giant flux creep model [20]. Our data are more consistent with the last model. The comparison of B*(T) for the two wires in Figure 9 indicates that a better pinning is associated to the multifilamentary wire, probably due to more defected grains induced by the reduced reaction time.

*4.3 Pinning Force*

In the framework of the assumed Kramer model, the IcB(B) curve describes the volume pinning forces acting on the fluxons at various B fields and $B_m = B^*(T)\, p/(p+q)$ is the field at which the force reaches a maximum.
In Figure 10 we compare the volume pinning force behaviour derived from our data. A limited shift of the maximum at higher field is observed for the monofilamentary wire. In any case the maximum force for these kind of wires is at B≤1.5 T, in agreement with the reported values for the wires prepared by an in situ process [4]. On the other hand, the ex situ process leads to a $B_m$ close to 4 T.

**5. Potential applications of the hollow wires**

The distinctive characteristics of the presented $MgB_2$ hollow wires are a very high density of the superconducting material and a very pure and regular polycrystalline microstructure, only the excess metallic magnesium being present as impurity, probably at the grain boundaries, which has no evident effect on the transport properties. Furthermore the presence of the central voids can be tailored to favour the wire refrigeration by a liquid coolant. In this respect the potential use of neon as coolant cannot be excluded. The transport properties of the wires at 25 K and around 2 T ($J_c \approx$ 300 A/mm$^2$) are compatible with many conventional power applications. The use of these wires in the higher magnetic field applications, in the range of 2÷4 T at intermediate lower temperatures, i.e. 15÷25 K, provided by the use of two-stage cryocoolers, could result in substantial energy saving with respect to the actual use of the liquid helium apparatus. Typical applications in these class of magnetic fields are the superconducting motors or generators, the medium-high field magnets for magnetic energy storage, more sensible MRI, magnetic separation, levitation systems and in general all the superconducting systems which require an user friendly refrigeration system.

Further advantages of the $MgB_2$ wires are a large availability of the raw materials and the ease of the manufacturig process, at least referring to the here presented techniques. These factors can bring to a wire high production rate, a condition that at this time cannot be reached by the competing BSCCO tapes. At this stage the main limitation in the expansion of the $MgB_2$ superconducting systems can be foreseen by the economical issues in the refrigeration technology.

## 6. Discussion and conclusions

The liquid infiltration process in obtaining $MgB_2$ compact superconducting part has been successfully applied also to the wire manufacturing , other than to bulk geometries. Several practical process features concurred to the reaching of the results here described. Among others we indicate the surprisingly high area reduction ratio of the magnesium rods, in a cold-working process when the rod is dressed by relatively hard powders, like the boron one; in this respect it is worth recalling that Mg usually presents a very limited ductility. Furthermore the high reaction rate that prevents any movement of the reacting powders from the zone they occupy after the mechanical processing of the wires. Regarding the materials here used in the lining and casing of the wires there are several possible alternative choices, in view of a better stabilization and an optimisation of the superconductor fill factor: the basic criterions to be maintained are the minimization of the interaction of these materials with the reacting elements and an enough mechanical strength of the casing that allows a good densification of the growing $MgB_2$ grains.

As far as the superconducting wire manufacturing is concerned the process presented here can be considered a substantial improvement with respect to the usual PIT process in term of easy of production and of the high performance of the superconductor. This does not exclude that a lot of problems remain to be solved for the industrial full exploitation of the $MgB_2$ wires. Among the most urgent problems we indicate the strain tolerance of the reacted wires , that, if solved, should determine the possibility to use the more friendly "react & wind" process to manufacture magnets. Other problems are related to the improvement of the transport characteristics: a) the enhancement of the pinning properties of the material in high magnetic fields, b) the limitation of the AC losses, c) the reaching of higher engineering critical current densities by a suitable superconducting wire design that combines the intrinsic high performance of the superconducting part with the need of an appropriate stabilization.

The last but not least issue for the $MgB_2$ exploitation is related to the development of more efficient and economical refrigeration systems , especially in the range of the higher power consumed at low temperature, i.e. >1 kW . The refrigeration system indeed is one the critical parts of the superconducting system in the competition where alternative classical systems are already working.

**Acknowledgements**
The Lecco Laboratory of CNR-IENI Institute is acknowledged for the support to the wires manufacturing. Claudio Orecchia is acknowledged for the SEM analysis

Table 1 – Area values of the two hollow wires (mm$^2$) (± 10%)

| Wire | SC Area | SC+Voids +Nb | Total | R=(SC+V+Nb)/Tot |
|---|---|---|---|---|
| Monofilament | 0.1 | 0.34 | 1.23 | 0.28 |
| 7-Filaments | 1.0 | 3.50 | 4.91 | 0.71 |

Table 2. Scaling parameters used to describe the critical current of the MgB$_2$ hollow wires

| Parameter | Monofilamentary wire | 7-Filaments wire |
|---|---|---|
| p | 0.5 | 0.8 |
| q | 5 | 5 |
| B*(0) (T) | 13.5 ± 0.5 | 11.39 ± 0.24 |
| α | 1.2 ±0.1 | 1.08±0.05 |
| C$_0$ (AT) | 28900 ± 2000 | 8830 ± 160 |
| β | 1.55 ± 0.37 | 1 |
| γ | 1.89 ± 0.45 | 0.82 ± 0.03 |
| T$_c$ (K) | 39 | 39 |

**Figure captions**

Figure 1 – Optical micrography of a cross section of the hollow monofilamentary wire

Figure 2 – Optical micrographs of cross sections of the 7-filament wire : a) precursor wire, b) annealed superconducting wire

Figure 3 – Backscattered electrons analysis of the interior of the hollow monofilamentary wire. The grey zone at the interface Nb/$MgB_2$ and some roughness in the $MgB_2$ part are the effects of interaction of the $MgB_2$ with the Nb sheet

Figure 4 - Morphology of the $MgB_2$ part in the monofilamentary wire

Figure 5 - Morphology of the $MgB_2$ part of the 7-Filaments wire

Figure 6 - SEM analysis of the Nb sheet and the adjacent $MgB_2$ parts in the 7-Filamnts wire

Figure 7 – Jc(B,T) of the monofilamentary wire

Figure 8 – Jc(B,T) of the 7-Filaments wire

Figure 9 – B*(T) calculated for the two hollow wires : (a) monofilamentary ; (b) 7-Filaments

Figure 10 – Field and temperature dependence of the volume pinning forces for the two hollow wires : a) monofilamentary , b) 7- Filaments

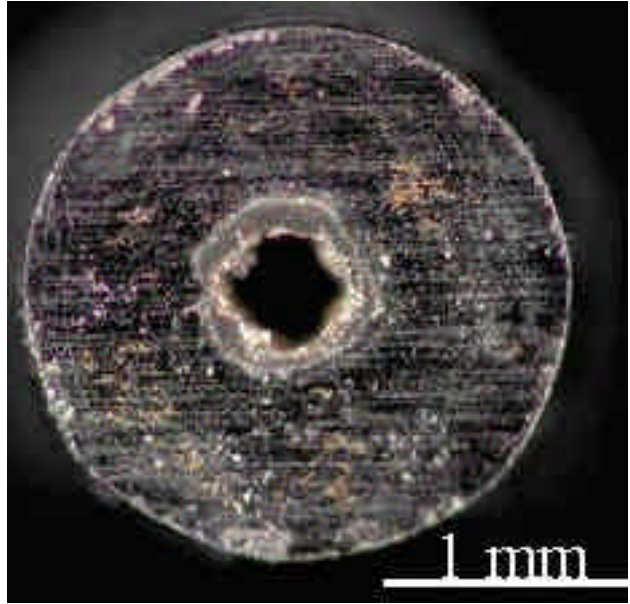

**Fig. 1**

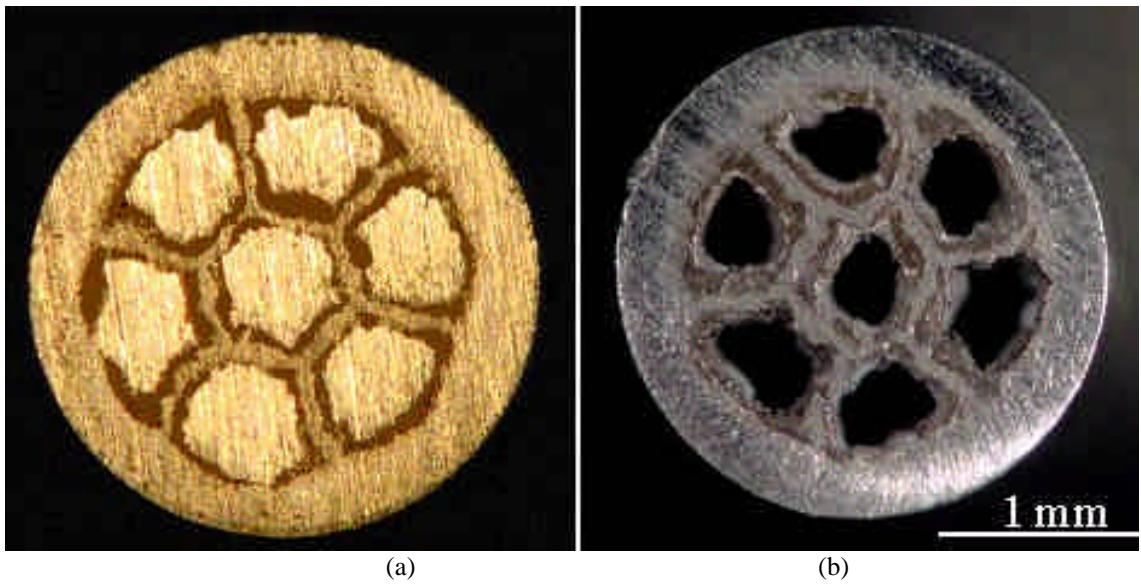

(a)            (b)

**Fig. 2**

Giunchi et al.

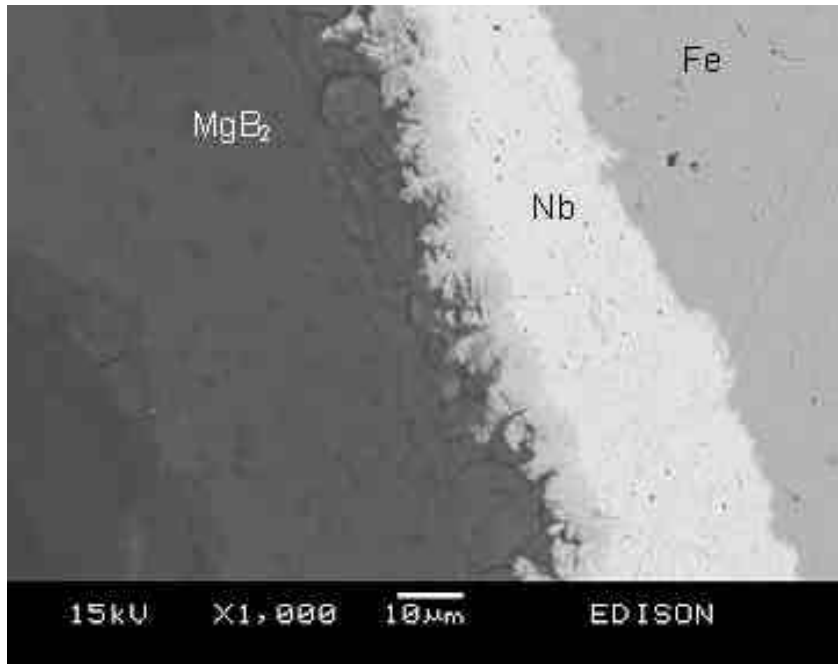

**Fig. 3**

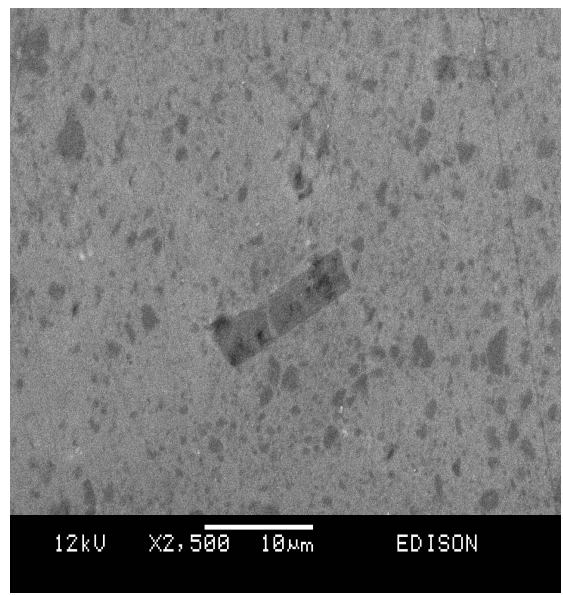

**Fig. 4**

Giunchi et al.

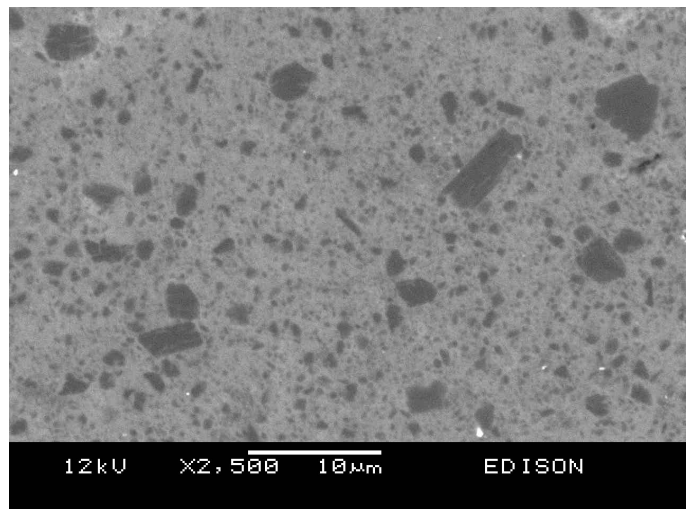

**Fig. 5**

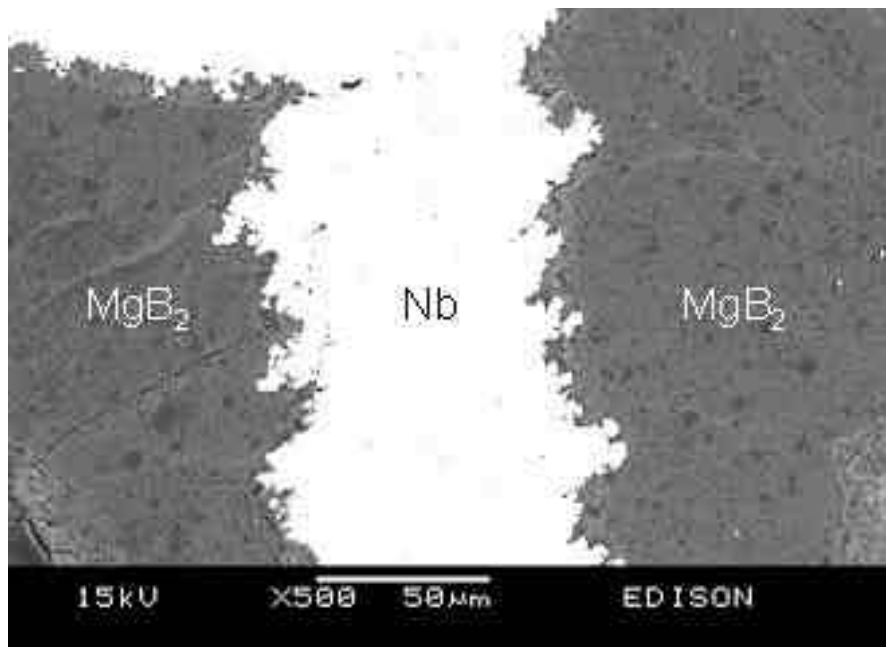

**Fig. 6**

Giunchi et al.

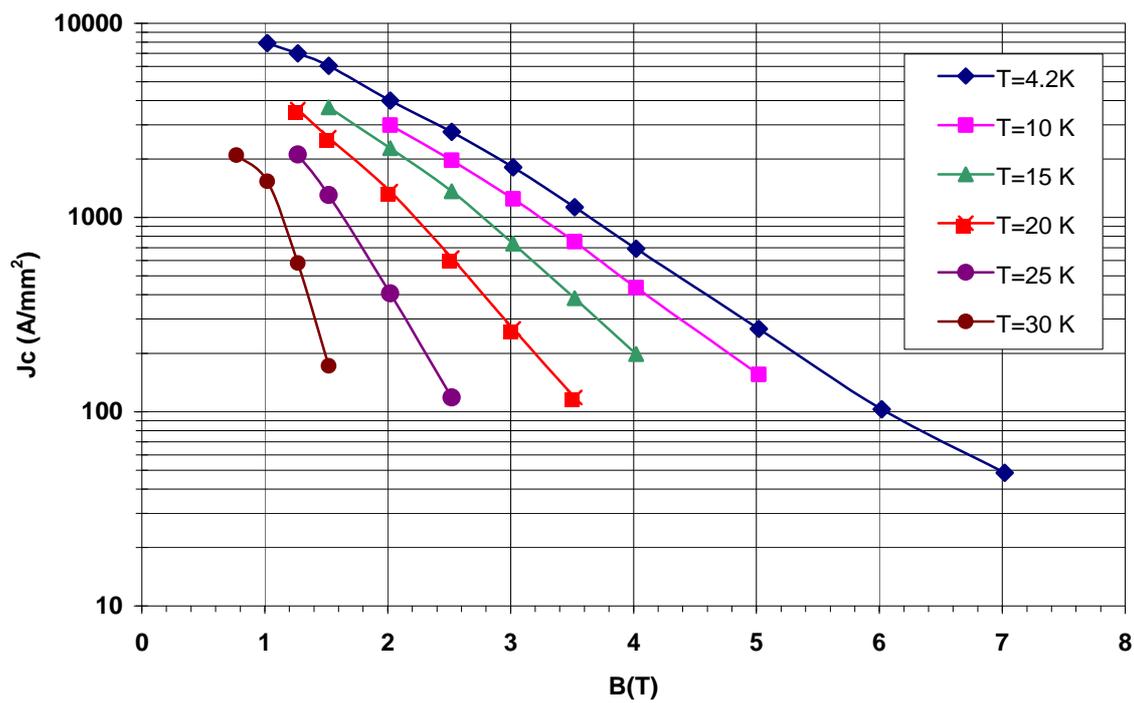

**Fig. 7**

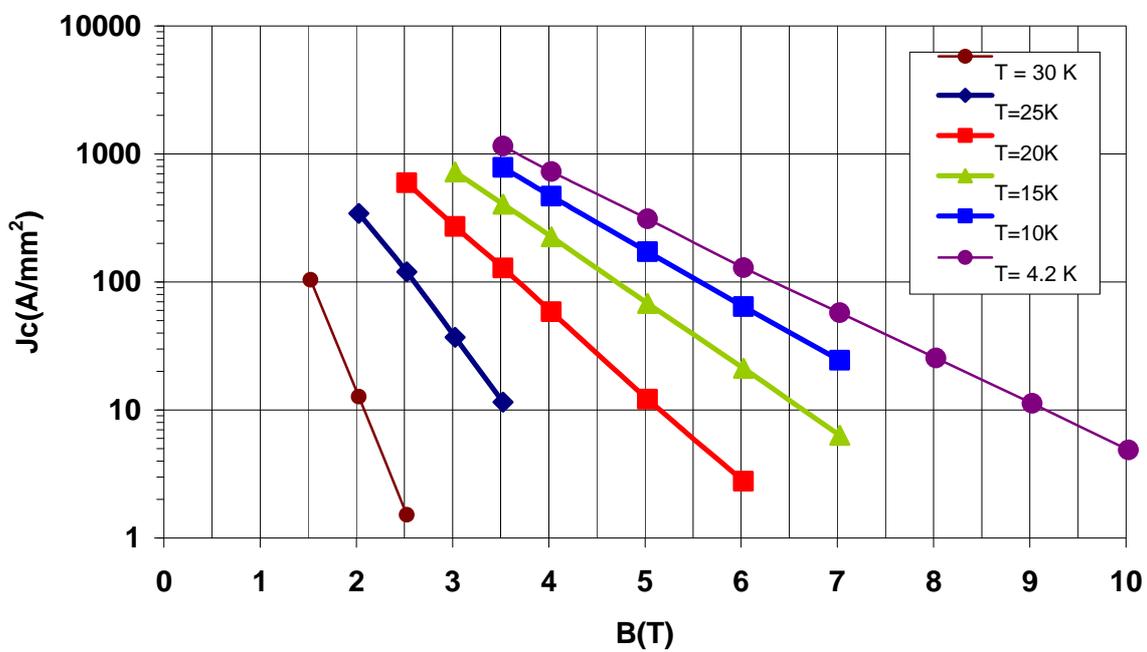

**Fig. 8**

Giunchi et al.

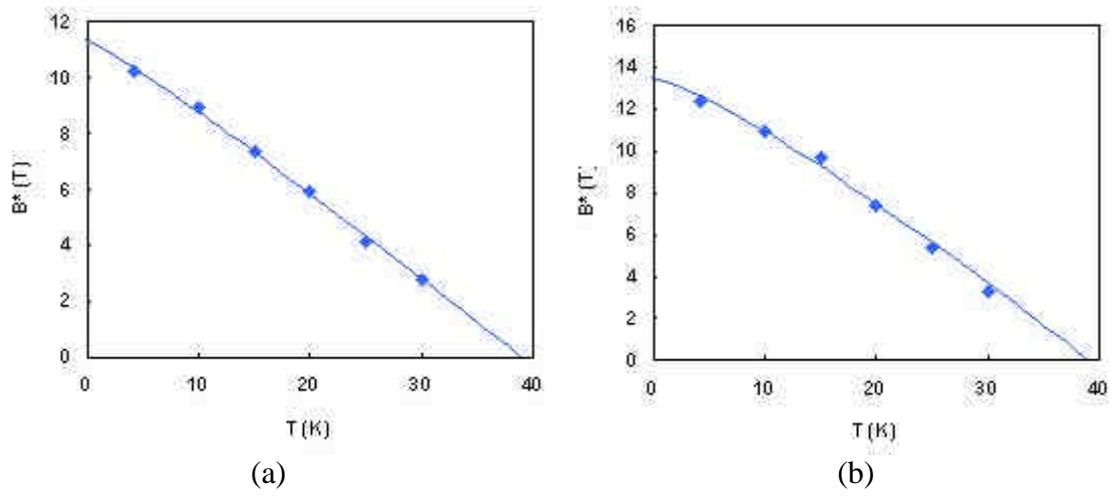

(a) (b)

**Fig. 9**

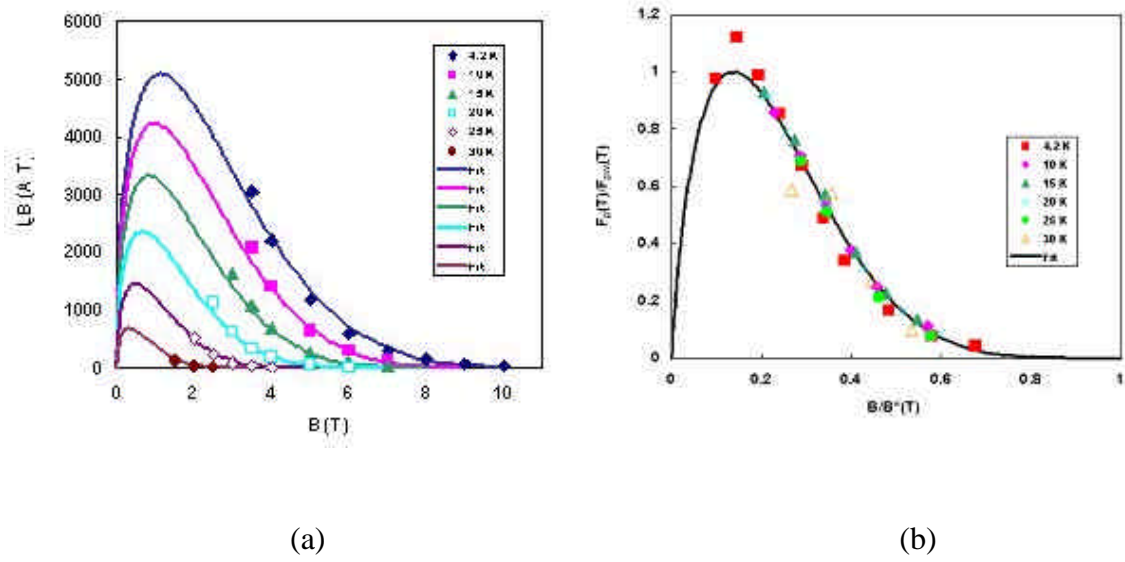

(a) (b)

**Fig. 10**

Giunchi et al.

14